\documentclass[utf8]{nf}  

\setcitestyle{square} 
\usepackage{url}
\usepackage{hyperref}
\usepackage{lineno,}
\usepackage{microtype,caption}
\usepackage{caption}
\usepackage{lscape}
\usepackage{multirow}
\usepackage{subfig}
\usepackage{graphicx}
\usepackage{amssymb}
\usepackage[export]{adjustbox}
\usepackage[onehalfspacing]{setspace}
\newcolumntype{L}{>{\raggedright\arraybackslash}X}
\usepackage{tabularx, booktabs}

\def\keyFont{\fontsize{8}{11}\helveticabold }
\def\firstAuthorLast{CUTE at SNOLAB} 
\def\Authors{
Philippe Camus\,$^{1,\dagger}$,
Jonathan Corbett\,$^{1}$,
Sean Crawford\,$^{1}$,
Koby Dering\,$^{1}$,
Eleanor Fascione\,$^{1,2}$,
Gilles Gerbier\,$^{1}$,
Richard Germond\,$^{1,2,\ddagger}$,
Muad Ghaith\,$^{1,\S}$,
Jeter Hall\,$^{3,4,5}$,
Ziqing Hong\,$^{4}$,
Andrew Kubik\,$^{3,4}$,
Adam Mayer\,$^{2,\#}$,
Serge Nagorny\,$^{1}$,
Payam Pakarha\,$^{1}$,
Wolfgang Rau\,$^{1,2,*}$, 
Silvia Scorza\,$^{3,4,5 \dagger\dagger}$ and
Ryan Underwood\,$^{1,2}$
}

\def\Address{
$^{1}$Queen's University, Kingston ON, Canada\\
$^{2}$TRIUMF, Vancouver BC, Canada\\
$^{3}$SNOLAB, Lively ON, Canada\\
$^{4}$University of Toronto, Toronto ON, Canada\\
$^{5}$Laurentian University, Sudbury ON, Canada
}

\def\corrAuthor{Wolfgang Rau; TRIUMF, 4004 Wesbrook Mall, Vancouver BC, V6T2A3}

\def\corrEmail{wrau@triumf.ca}

\usepackage[dvipsnames]{xcolor}

\begin{document}

\onecolumn
\firstpage{1}

\title {The Cryogenic Underground TEst (CUTE) Facility at SNOLAB} 
\author[\firstAuthorLast ]{\Authors} %This field will be automatically populated
\address{} %This field will be automatically populated
\correspondance{} %This field will be automatically populated
\extraAuth{New address\\
$^{\dagger}$Institut N\'eel, CNRS, Grenoble, France\\
$^{\ddagger}$Institute for Quantum Computing, University of Waterloo, Waterloo ON, Canada\\
$^{\S}$Zayed University, Dubai, United Arab Emirates\\
$^{\#}$Lancaster University, Lancaster, United Kingdom\\
$^{\dagger\dagger}$Univ.\ Grenoble Alpes, CNRS, Grenoble INP*, LPSC-IN2P3, Grenoble, France
}% If there are more than 1 corresponding author, comment this line and uncomment the next one.
%\extraAuth{corresponding Author2 \\ Laboratory X2, Institute X2, Department X2, Organization X2, Street X2, City X2 , State XX2 (only USA, Canada and Australia), Zip Code2, X2 Country X2, email2@uni2.edu}
\maketitle
\begin{abstract}
\section{}
Low-temperature cryogenics open the door for a range of interesting technologies based on features like superconductivity and superfluidity, low-temperature phase transitions or the low heat capacity of non-metals in the milli-Kelvin range, that may be employed by a variety of applications. Devices based on these technologies are often sensitive to small energy depositions as can be caused by environmental radiation. The Cryogenic Underground TEst facility (CUTE) at SNOLAB is a platform for testing and operating cryogenic devices in a low-radiation background environment. Reaching a base temperature of $\sim$\,12$\;$mK, the large experimental chamber ($\mathcal{O}$(10)$\;$L) can hold a payload of up to $\sim$\,20$\;$kg and provides a radiogenic background event rate as low as a few events/kg/keV/day in the energy range below about 100$\;$keV, and a negligible muon rate ($\mathcal{O}$(1)/month). It was designed and built in the context of the Super Cryogenic Dark Matter Search experiment (SuperCDMS) that uses cryogenic detectors to search for interactions of dark matter particles with ordinary matter and has been used to test SuperCDMS detectors since its commissioning in 2019. In 2021 was handed over to SNOLAB to become a SNOLAB user facility after the completion of the testing of detectors for SuperCDMS. The facility will be made available for projects that benefit from or require these special conditions, based on proposals assessed for their scientific and technological merits. This article describes the main design features and operating parameters of CUTE.

\tiny
 \keyFont{ \section{Keywords:} Cryogenic, Low Radiation Background, Underground Science, SNOLAB, Dark Matter, Rare Event Searches}
 %All article types: you may provide up to 8 keywords; at least 5 are mandatory.
\end{abstract}

\section{Introduction}
\label{sec:intro}
Cryogenic particle detectors are among the detectors with the best energy resolution, reaching eV-scale thresholds (see e.g. \cite{CRESST:2020, ECHo:2023, PD2:2021, NUCLEUS:2020}). When combined with intrinsic amplification, this is even possible in large devices (up to kg scale) \cite{CDMSliteR2L:2018, EDELWEISS:2023} or goes well below an eV in smaller devices \cite{HVeV:2020}. Hence, this is often the technology of choice for detecting low-energy interactions. If the particle interaction under investigation is also a rare process, a low-background facility is required to conduct the respective experiment. In the case of a low-threshold detector with a large mass, it may be necessary to operate them in a low-background environment just to study the basic detector performance: a low threshold combined with a large mass typically leads to a high rate of background events; if the response time of the detector is slow (which is often the case in cryogenic detectors), the detector would be overwhelmed with interactions if not properly shielded. Finally, if a very low background is required for the experiment, cosmogenic activation could render a detector useless if it was tested extensively without protection from cosmogenic radiation.

The Super Cryogenic Dark Matter Search (SuperCDMS) \cite{SuperCDMS:2022kse}, presently under construction at SNOLAB, a deep underground laboratory hosted inside the active Vale Creighton mine near Sudbury, ON, Canada, combines all the above requirements: massive low-threshold detectors for the search of very rare interactions of dark matter particles, made in part of -- and surrounded by -- materials that get easily activated by cosmogenic radiation~\cite{SuperCDMS:2018tqu}. This motivated the design and construction of a Cryogenic Underground TEst facility (CUTE) at SNOLAB~\cite{Camus:2018, Rau:2020}, to enable the testing of the new SuperCDMS detectors under low-background conditions, without the danger of cosmogenic activation.

However, from the beginning, consideration was given to a possible use of the facility after the primary goal of testing SuperCDMS detectors would be concluded. This informed a number of design choices that now make this a convenient facility for testing and operating not only particle detectors, but also other devices with low operating temperature that benefit from (or require) very low levels of background radiation. In 2021, SNOLAB has taken on the responsibility of maintaining and upgrading the facility and will be making it available to new users based on the merit of their proposals. 

In this paper, we introduce the design considerations for the facility and describe its different components and subsystems and their performance, before ending with a short discussion of possible uses beyond SuperCDMS that might benefit from the special conditions provided by CUTE. While the focus is on the facility as such, we will discuss some of the specific solutions implemented for SuperCDMS as examples where appropriate.

\section{Design Requirements and Considerations}
\label{sec:design}
The original motivation for the CUTE facility was the testing of the new SuperCDMS detectors~\cite{SuperCDMS:2016wui}. Besides understanding basic detector parameters, a number of calibration measurements are planned, including a measurement of the neutron-gamma discrimination power of the detectors to better than 10\textsuperscript{-6} (meaning that less than one in 10\textsuperscript{6} gamma interactions is misidentified as neutron interaction). The minimal requirements the facility would have to fulfill are defined by the needs for operating the detectors and the necessity that the calibration measurements can be conducted without major interference from background radiation. 

The SuperCDMS detectors consist of cylindrical Ge or Si crystals, 10$\;$cm in diameter and 3.3$\;$cm thick (1.4 and 0.6$\;$kg respectively), equipped with superconducting transition-edge sensors (TESs) that require an operational temperature below about 30$\;$mK.\cite{SuperCDMS:2016wui} That implies that a dilution refrigerator is needed for extended operation. The detectors come in two denominations: iZIP detectors where both phonons and charges are measured for optimal background identification and discrimination, and high-voltage or HV detectors that are operated with a bias voltage of order of 100$\;$V (compared to just a few V for the iZIPs) where the Neganov-Trofimov-Luke effect \cite{Neganov:1985,Luke:1988} leads to a large increase of the phonon-signal and in turn a very low effective energy threshold. The detectors are arranged in stacks of six, attached to a structure that provides the mechanical, thermal and electrical connection and includes the central elements of the first-stage amplifiers. The complete assembly has a mass of up to $\sim$\,20$\;$kg and is referred to as {\it tower}, where the tower without the detectors is called the {\it tower body} and the electronic elements attached to the tower are referred to as {\it cold electronics}. The sensors and some elements of the cold electronics require magnetic shielding (ideally $\lesssim$\,1$\;\mu$T). 

For calibrating these detectors, the maximum beneficial rate is about 7$\;$Hz before being limited by pile-up, so the overall background rate would need to be $\ll$1$\;$Hz to ensure a good signal-to-background ratio during calibrations. For the neutron-gamma discrimination measurement, the neutron rate would need to be of order of one neutron interaction per day per detector or less in the energy range of interest ($\sim\;$1-50$\;$keV), since the maximum number of gamma interactions that can be accumulated per day is a few times 10\textsuperscript{6}.

To achieve the background requirements, the facility must be located underground to reduce the cosmic-ray induced neutron flux \cite{Mei:2006}. As previously discussed, a significantly reduced cosmic ray flux is also necessary to protect the detectors from cosmogenic activation. Additional shielding is necessary to reduce the environmental radiogenic neutron and gamma flux~\cite{SNOLABhandbook}, and some effort is required to avoid introducing contaminants into the experimental setup. A decision was taken early on in the planning process to take additional steps towards a considerably lower background level. The additional effort to accomplish this was modest and it gives the best perspective for a useful life of the facility after the completion of the SuperCDMS tests.

Other considerations include the strong susceptibility of the SuperCDMS detectors to mechanical vibration, due to their large mass, and the susceptibility of the readout electronics to electromagnetic interference. In addition to the right environment for the measurement, it is essential that the installation of the detectors into the facility does not significantly increase their exposure to contaminants such as dust or radon. Hence, a dedicated cleanroom with low-radon air supply that can host the cryostat for payload changes is an important part of the facility. 

\section{Cryogenics}
\label{sec:cryogenics}
The CUTE cryostat is cooled by a cryogen-free (dry) dilution refrigerator from CryoConcept, with a base temperature of approximately 12$\;$mK. The cryostat has six thermal stages with nominal temperatures of 300$\;$K, 50$\;$K, 4$\;$K, 1$\;$K, 100$\;$mK, and 10$\;$mK, respectively. The lowest three stages are referred to as the {\it Still} (ST), {\it Cold Plate} (CP) and {\it Mixing Chamber} (MC) stages, after the respective functional components of the dilution refrigerator while the other two cold stages are referred to as 4K and 50K stages. Mechanical connections between the stages are made by stainless steel or G10 standoffs which provide solid structural connections with low thermal conductivity. The room-temperature stage consists of a stainless-steel vacuum vessel referred to as the {\it Outer Vacuum Can} (OVC). Copper cans acting as thermal radiation shields are mounted on the 50K, 4K, and Still stages; the 50K-can is wrapped in multiple layers of aluminized Mylar super-insulation to reduce the thermal load from the OVC. A separate experimental stage is thermally anchored to the MC stage with three copper bars, providing a large volume ($\sim$\,25$\;$cm diameter and 30$\;$cm high) in which the experimental payload can be mounted.

The 50K and 4K stages are cooled by a pulse-tube cryocooler (PTC). In order to minimize the coupling of vibrations from the PTC into the cryostat, CryoConcept developed a technique that avoids mechanical connections between the cold-stages of the PTC and the cryostat, called the Ultra-Quiet Technology\textsuperscript{TM} (UQT). Thermal contact is provided by a gas link: the PTC's cold head is installed inside the Still pumping line of the DU. Gold-plated copper disks with a concentric ring structure mounted on the cold head stages are interleaved with corresponding disks thermally connected to the 50K and 4K stages of the cryostat, providing a large effective heat transfer surface while maintaining a compact design and a narrow gas gap ($\sim$\,1$\;$mm) as the circulating gas meanders through the ring structure.

The lower thermal stages of the cryostat are cooled by the dilution unit (DU). The cool-down process includes two distinct steps. During the precooling, helium is circulated past the PTC before infusing the DU while bypassing the main impedance. This step by which the DU is cooled to $\lesssim$\,4$\;$K takes roughly three days. In the second step, the \textsuperscript{3}He/\textsuperscript{4}He-mixture is condensed in to the refrigerator, starting the main cooling cycle. The base temperature (11-12$\;$mK without any payload) is reached about 8$\;$h after the start of the condensing step. With a payload installed it is typically observed that the temperature settles at around 15$\;$mK, before slowly (over 1-2~weeks) cooling down to $\sim$\,12$\;$mK (owed likely to not, or imperfectly, annealed Cu parts in the setup). The cooling power at 100$\;$mK is $\sim$\,200$\;\mu$W. The warm-up to room temperature takes nearly a week if the cryostat is left under vacuum; however, a system was installed that allows for the introduction of nitrogen gas into the OVC which can reduce the warm-up time to less than three days.

The SuperCDMS detector tower body includes three thermal stages mirroring the three lower stages of the cryostat (MC, CP and ST). This allows the cold electronics and the wiring between the MC and ST stages to be mounted directly to the tower. The main weight-bearing mechanical connection is at the lowest temperature stage which is attached to the experimental stage of the cryostat, with the detectors sitting above the stage and the tower body with the warmer thermal stages sitting below. For shielding of the detectors against infrared radiation (IR) from higher temperature stages, a copper can (referred to as {\it top hat}) surrounding the detector stack is mounted on the experimental stage. To minimize radioactive background it is made entirely from copper, held together and connected to the experimental stage by a total of only four screws; brazing or soldering was avoided as these techniques known to add noticeable levels of radio-contaminants. 

The thermal connection to the CP stage of the tower is provided by a braided and annealed copper heat sink wire attached to a long copper bar which reaches from the CP stage of the cryostat to below the experimental stage. The Still stage of the tower is thermally connected to the bottom of the Still can by means of a $\sim$\,0.3$\;$mm thick copper membrane that also closes the Still can volume to avoid thermal radiation from the 4K stage to reach the MC stage. In order to avoid stress on the tower due to differential thermal contraction in the vertical direction between the tower and the cryostat, the membrane has an undulation half way between the tower and the rim of the can which allows it to compensate for a few mm of relative vertical movement. 

An extension to the 4K can was designed and built out of copper and held together by screws only, again to avoid brazing and the related increase in radioactive background. This extension has a removable bottom lid and includes feedthroughs for the six readout cables for the SuperCDMS towers. These cables are guided up along the outside of the 4K can where the feedthrough and the connection to the can provide the heat sinking to that stage; they are further heat-sunk at the 50K stage before they are connected to the room temperature vacuum feedthrough.

\begin{figure}
    \centering
    \includegraphics[width=0.8\textwidth]{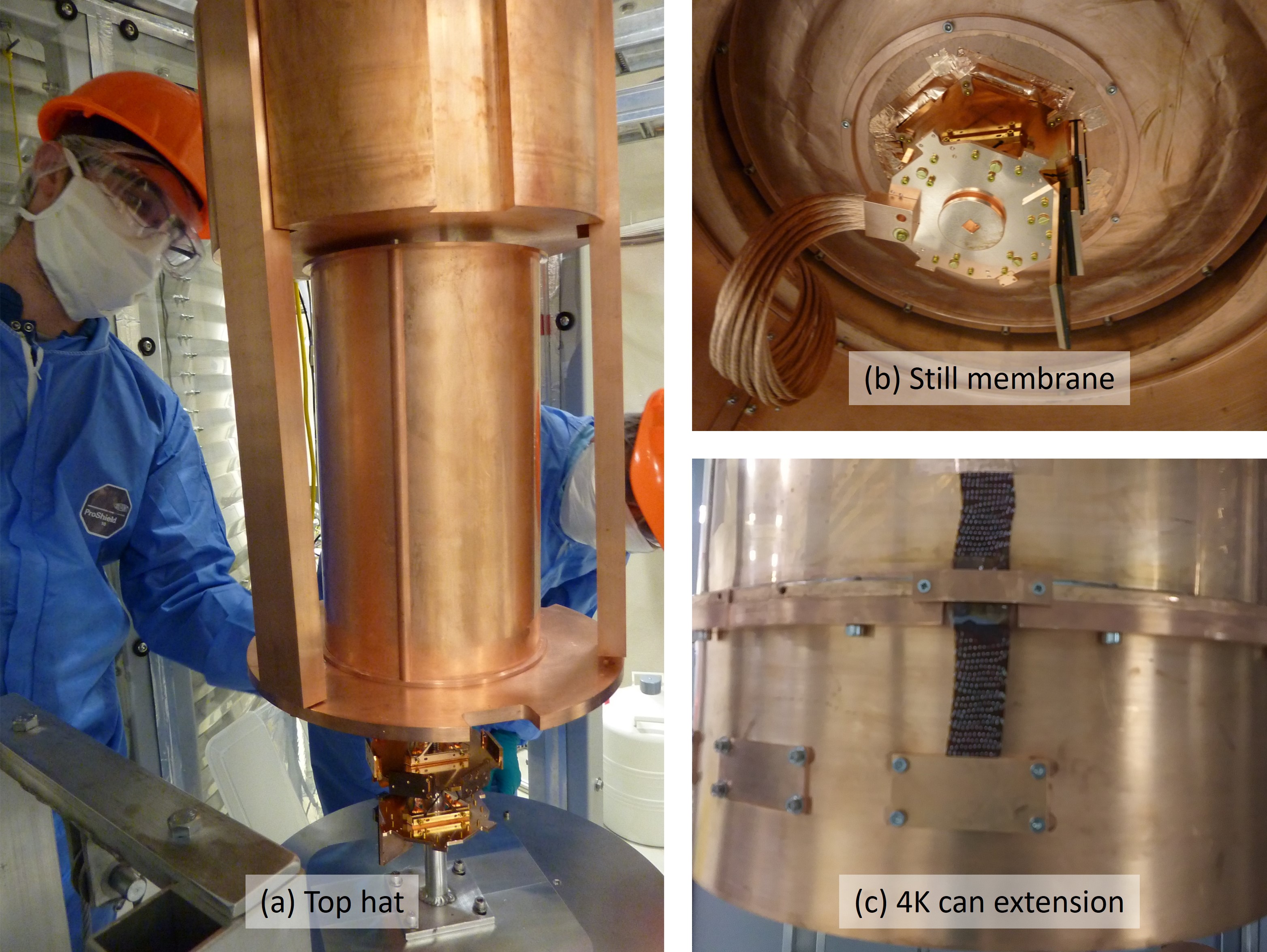}
    \caption{(a) The IR shield at the MC stage ({\it top hat}) is mounted on the experimental stage which in turn is attached to the cryostat by the three MC bars. Directly above the top hat is the internal lead shield (see Sec.\ \ref{sec:background} for more detail on the shield). The half-pipe shaped walls of the top hat sit in grooves in its top plate, the bottom ring and the connecting rods, and the whole assembly is held together with the two screws connecting the top plate to the rods and two screws that attach the top hat to the experimental stage (connecting through the bottom ring to the rods). (b) The Still membrane with the visible circular undulation is mounted, connecting the bottom of the Still can to the Still stage of the SuperCDMS tower, acting as thermal link as well as closing off the Still volume to IR from the warmer thermal stages. (c) The 4K extension, consisting of a wall section with screwed on flange and bottom plate. When needed, the cover plates in the walls are replaced by the easy to install IR blocking cable feedthroughs (also shown in the picture).}
    \label{fig:cryo_1}
\end{figure}

Figure \ref{fig:cryo_1} shows pictures of the MC top hat together with the experimental stage mounted on the MC bars, the Still membrane, and the 4K extension with the IR blocking cable feedthroughs.

The dilution refrigerator was delivered with the thermometry required for its operation and an additional unused 12-pin vacuum feedthrough installed. This was utilized for custom wiring for three additional thermometers. The wiring is routed to the MC stage of the cryostat and heat-sunk at the different thermal stages using custom-designed printed circuit boards that also act as IR blocking feedthroughs. 4-pin connectors at each feedthrough allow for easy connection of auxiliary thermometry at the desired thermal stage. In addition, a coaxial wiring solution for the future operation of sensors that require the transmission of high-frequency signals has been developed. An initial test showed that this wiring does not impact the cryogenic performance.

For payload changes, the cryostat needs to be moved from its operating location inside the shielding to the facility's cleanroom which requires all electrical as well as gas and vacuum connections to be removed. The high-pressure helium hoses for the PTC are connected to the rotary valve using quick-connect style connections that automatically seal when disconnecting and minimize air inclusion when connecting. When the cryostat is placed inside the shielding, many of the other standard connection points are difficult to reach. Hence, all connections for the helium mixture and the cryostat vacuum were routed to a conveniently located {\it connection plate} which has manual valves to close off all the lines going to the cryostat. All lines that may get cold during operation (primarily during the precooling stage) are insulated to minimize frost production and the subsequent dripping water after warming up. A multi-connector is used for conveniently connecting and disconnecting simultaneously all air-pressure lines for the pneumatic valves.

\begin{figure}
    \centering
    \includegraphics[width = 0.70\textwidth]{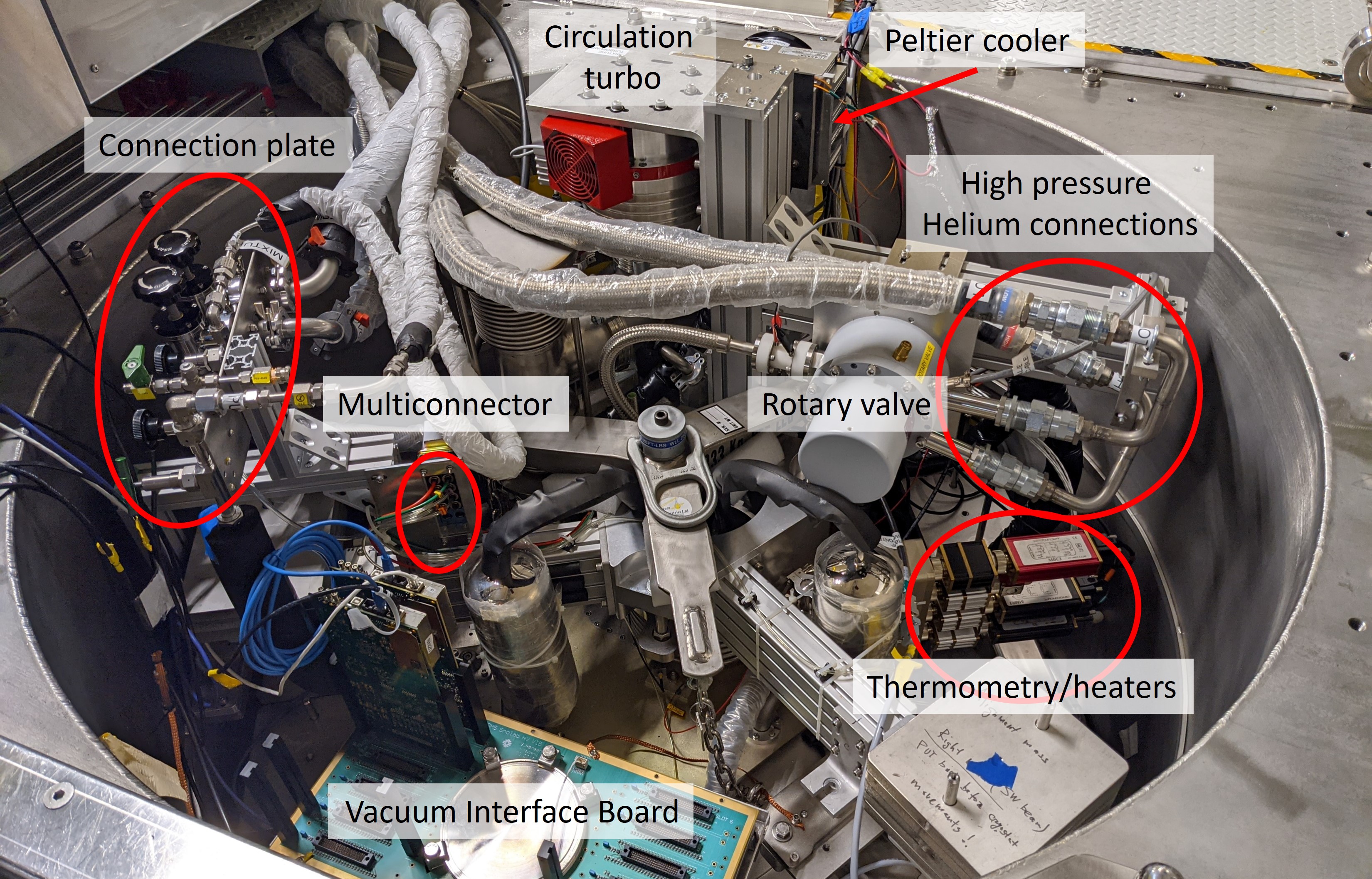}
    \caption{Top view of the cryostat when installed and fully connected in its operating location inside the drywell. Labelled in the picture are on the left the connection plate collecting all vacuum and helium-mixture lines that need to be disconnected for moving the fridge. Left of the centre is the multiconnector for the pressurized air lines for the pneumatic valves. In the top centre the turbo pump for the helium mixture is visible, and attached behind it is the Peltier cooler. To the right of the centre is the rotary valve and next to it the connection points for the high-pressure helium lines; next to those are the thermometry and heater connections. At the bottom centre of the picture is the Vacuum Interface Board (VIB) for connecting the SuperCDMS readout electronics.}
    \label{fig:cryo_2}
\end{figure}

Part of the circulation system for the helium mixture is a turbo molecular pump mounted on top of the refrigerator which requires cooling. Due to the necessity of regular disconnections, the use of cooling water would be inconvenient and bring a non-negligible risk of water leaks. Therefore, a Peltier cooler is used to cool the turbo pump. The temperature is regulated by a feedback control system. The original control circuit used medium-frequency switching of a high power line, leading to significant electromagnetic emission that can be picked up by the readout electronics. To mitigate this effect an alternate control was implemented that instead uses voltage control. The power supply is a high-frequency switching power supply where the switching frequency is above the bandwidth of the presently used readout electronics, thereby removing the electromagnetic interference introduced by the original control circuit. If a future use of the facility is also susceptible to the interference from the high-frequency switching power supply, it can easily be replaced by a non-switching power supply. Figure \ref{fig:cryo_2} shows a top-view of the refrigerator when fully connected at its operating location inside a drywell in the centre of the shielding water tank (for more details of the facility layout see \ref{sec:background}) .

\section{Vibration Isolation}
\label{sec:vibration}
\begin{figure}
    \centering
    \includegraphics[width=1\textwidth]{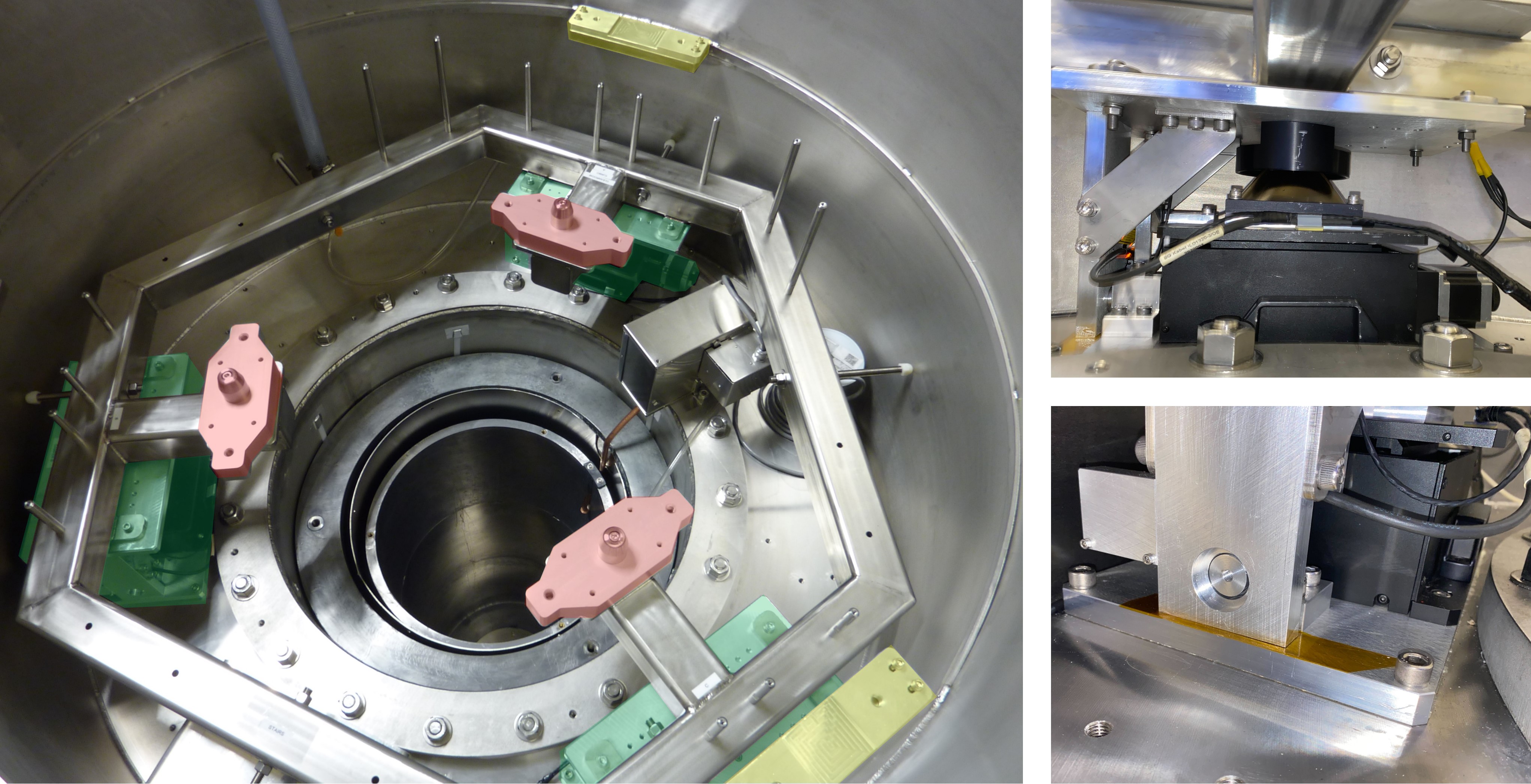}
    \caption{Left: The suspension system's steel frame with the mounting points for the cryostat (red highlights) sits on three dampers (damper assemblies highlighted in green) to mechanically isolate the cryostat from the vibrations transmitted through the drywell. Also visible are two of the mount points for the upper part of the refrigerator on the drywell (yellow highlights), as well as the shielding inside the lower part of the drywell (outer and inner layers of lead with a thin magnetic shield in between; see Sec.~\ref{sec:background}). Located on the top right in the left picture is the gamma calibration system (see Sec.~\ref{sec:calibration}). Top right: one of the three damper assemblies. A labjack (black) driven by a stepper motor (to the right of the labjack) sits on the drywell; the damper (elastomer cup) is mounted on the movable stage of the labjack and carries the suspension frame. The positions of both, the movable labjack stage and the suspension frame are measured relative to the drywell. Visible to the left of the labjack is a vertical aluminum bar rigidly attached to the suspension frame. Right bottom: The vertical aluminum bar has a circular cutout. A rod attached to the drywell penetrates this cutout with a nominal clearance of 1.0$\;$mm all around, limiting the movement of the suspension frame relative to the drywell.}
    \label{fig:suspension}
\end{figure}

Several strategies are utilized to mitigate the transmission of vibrations from the environment and the cryogenic equipment mounted on top of the refrigerator to the experimental stage. Typically, the pulse tube cooler's rotary valve is the largest source of vibrations in a dry dilution refrigerator. The refrigerator used at CUTE features a double-frame design to effectively decouple the top with the rotary valve and the turbo pump from the lower part where the detectors are installed. The upper part is rigidly connected to the drywell, while the lower part is connected to a stainless steel frame, the {\it suspension frame}, which sits on three soft elastomer cup dampers (Newport ND20-A) as shown in Fig.~\ref{fig:suspension}. The upper and lower parts of the refrigerator are connected only by the outer wall of the Still pumping line in form of a bellows with very low stiffness (16$\;$N/mm along its axis and 167$\;$N/mm laterally), effectively decoupling the two parts. The UQT discussed in Sec.~\ref{sec:cryogenics} further minimizes the transmission of vibrations from the rotary valve of the PTC to the lower part of the cryostat.

With no stiff mechanical connection between the two parts, the low pressure inside the pumping line together with its large diameter leads to a strong upwards force from the atmospheric pressure in the lab acting on the lower (floating) part of the refrigerator which would compress the soft bellows. This is counteracted by extra weights placed on the suspension frame. 

Due in part to the design of the ventilation system at SNOLAB, the pressure in the lab can fluctuate by up to $\sim$\,20$\;$\% over the course of a few hours. The resulting time-varying force would lead to a change in position of the cryostat relative to the top part of the refrigerator given by the stiffness of the dampers ($\sim$\,80$\;$N/mm). This would change the cross section of the gas gap and hence impact the thermal profile of the cryostat: an increased gas gap would improve the Still pumping efficiency, cooling it down, but reduce the thermal link between cold head and the 4K and 50K stages of the cryostat. To compensate for the pressure variation, each of the three dampers supporting the suspension frame is mounted on the movable stage of a labjack. These stages are driven by stepper motors operated by a micro-controller. Given the relatively slow changes of the environmental pressure, a response time of order of seconds is by far fast enough, and so a software-based feedback system is implemented. Mounted to the suspension system, on one side of each damper assembly, is a vertical aluminum bar with a cutout for a rod that is mounted on the drywell. The cutout is only marginally larger than the rod's cross section, allowing for a maximal deviation of the actual from the nominal position of the suspension frame of one millimeter (vertically and laterally). Such a tight tolerance is necessary to avoid a touch inside the cryostat between the cold stages of the PTC and the cryostat stages. In the lowest possible position, the three aluminum bars carry the full weight of suspension frame and cryostat which then are no longer floating. To monitor the position of the suspension system and the state of the dampers, each damper assembly has an optical sensor and a Hall effect sensor. The optical sensors track the distance between the suspension frame and the drywell and thus the relative position of the two parts of the refrigerator; the Hall effect sensors are used to measure the position of the movable stage of the labjack. Together with the optical sensors, they are used to determine the compression of the dampers. This information is important as the performance of the dampers depends on their compression. More information about the control algorithm of the suspension system is presented in Sec.~\ref{sec:slow-control}.

\begin{figure}
    \centering
    \includegraphics[width=0.85\textwidth]{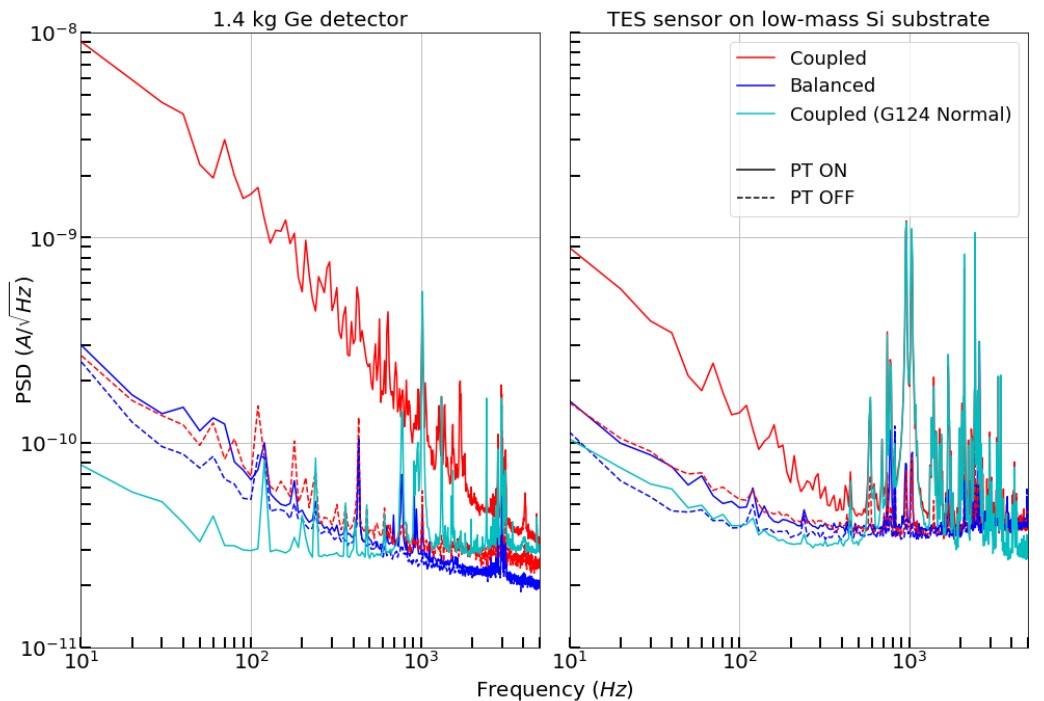}
    \caption{Noise power spectra from a 1.4$\;$kg germanium detector (G124) and a small silicon substrate with a transition edge sensor under various conditions of the experimental setup (figure from \cite{Germond:2023}). Operating conditions include PTC (PT in the legend, our main source of vibrations) on and off with the suspension system in its normal operating mode ({\it balanced}) as well as tilted to introduce a mechanical (as well as electrical) connection between the PTC cold head and the cryostat ({\it coupled}), the latter with G124 sensitive (sensor in its superconducting transition) as well as insensitive (sensor in normal conducting state) but no change in the readout conditions of the small device. There is a drastic increase in noise in G124 in the coupled state with the PTC on while no significant increase is seen in the small sample as long as G124 is insensitive, showing that the higher noise in the small sample comes from cross talk in the readout system. These findings support the hypothesis that the noise is indeed coupled through vibrations rather than electrically (the latter would have affected the small devices as much or more than the big one) and thus show the effectiveness of the suspension system.}
    \label{fig:g124_vib}
\end{figure}

The performance of the vibration isolation was verified using a prototype 1.4$\;$kg, germanium SuperCDMS HV detector (referred to as G124), as these large sized detectors have been observed to be sensitive to vibration-induced noise. To separate vibrational noise from noise caused by electromagnetic interference, a transition-edge sensor (similar to the TESs of the SuperCDMS detectors) on a small silicon substrate (chip) was operated alongside G124. This device is far less sensitive to vibrational noise but more sensitive to electromagnetic interference. For these tests, the state of the suspension system was changed between its normal (balanced) configuration, and a coupled configuration where it was intentionally tilted to introduce a contact between the cold stage of the PTC and the cryostat. The presence and absence of the touch was verified by checking the electrical continuity between the pulse tube's cold head and the cryostat which are normally electrically isolated. For each configuration, data from the detectors were collected with the PTC running and switched off. Figure \ref{fig:g124_vib} shows the noise power spectra of these tests for both G124 and the TES sensor on the chip. While in the coupled state there is a noise difference at low frequency of almost a factor of 40 between PTC on and off compared to only a very small difference (order of 10-20$\;$\%) in the balanced state. The noticeable difference between PTC on and off in the coupled state for the TES chip can be traced back to cross-talk in the readout between the two devices and is not related to actual noise in the TES chip (the excess disappeared when the current in the sensor on G124 was increased until the sensor went into its normal-conducting and thus insensitive state while no change was applied to the TES chip). The fact that the PTC-induced noise couples so strongly to G124 and only weakly to the TES chip confirms that the majority of the effect is indeed caused by mechanical vibration; all other effects would be as strong if not stronger in the smaller device compared to G124. The fact that the excess noise essentially disappears when the suspension system is balanced demonstrates the effectiveness of the system, even though this test is not suited to quantify the level of vibration reduction or residual vibrations.

For a more direct measurement of vibrations, a triaxial and a single-axis accelerometer are available. However, they are presently mounted on top of the cryostat (not inside), show a relatively high electronic noise level and there is indication that they pick up acoustic noise. Hence, some improvements will have to be implemented in order to use them for sensitive diagnostics. However, they were used to demonstrate that a high flow of purge gas in the drywell (the purge gas is used to avoid radon-rich air from the lab to enter the space; see Sec.~\ref{sec:background}) causes the cryostat to vibrate. The purge gas flow was optimized to a level where the vibrations caused are negligible (assessed by measurements with the accelerometer as well as a detector) while the radon level in the drywell stayed low. \cite{Corbett:2021}

\section{Shielding and Background}
\label{sec:background}
Background radiation comes from various sources; natural radioactive contamination, mainly potassium ($^{40}$K) and the uranium (U) and thorium (Th) decay chains, in the environment and the experimental setup itself produce high-energy gamma-rays as well as neutrons via ($\alpha$,n)-reactions and spontaneous fission. In the immediate vicinity of the sensitive components of the experiment, short-range radiation such as alphas, betas, and low-energy gammas need also to be considered. Lastly, cosmic radiation contributes muons, high-energy neutrons generated by muon interactions, and hadronic showers. 

The approximately 2$\;$km of rock-overburden at SNOLAB completely remove the hadronic component of the cosmic radiation and reduce the cosmic-ray muon flux by several orders of magnitude to $<$\,0.27$\;\mu$/m$^{2}$/day \cite{SNOLAB}, and with it the high-energy neutron flux. The experimental space inside the CUTE cryostat is further protected by layers of passive shielding to absorb or moderate the environmental radiation. Finally, some effort was made to minimize the level of contaminants inside the facility itself.

The CUTE shielding was designed based on Monte Carlo simulations considering the attenuation of external radiation as well as the acceptable levels of contamination of the shielding materials. The outer layer of shielding consists of a water tank with a stainless steel drywell in the centre to host the cryostat. The water tank filled with Ultra Pure Water (UPW) from the SNOLAB water purification plant provides a shielding thickness of $\sim$\,1.5$\;$m on the side and $\sim$\,1$\;$m  at the bottom, reducing the external gamma radiation by about a factor of 200~\cite{Liu:2011}. To avoid biological growth in the water tank, the UPW is  circulated in regular intervals through a container with bromine tablets. This is scheduled for times when no measurements are taken with the facility to prevent possible electromagnetic interference caused by the water circulation pump. A deck structure holds the drywell in place and provides access to the top of the cryostat.

Inside the drywell, surrounding the cryostat, are two layers of lead, the outer layer (8.7$\;$cm on the sides and 13$\;$cm on the bottom) with low activity of \textsuperscript{210}Pb, and the inner layer (2$\;$cm sides and bottom) with very low activity of \textsuperscript{210}Pb\footnote{``Faible Activite'' (FA) lead with about 40$\;$Bq/kg, and ``Tres Faible Activite'' (TFA) lead with about 6$\;$Bq/kg of  $^{210}$Pb, from the Fonderie de Gentilly, France}. Located between the two layers of lead is a $\mu$-metal shield (from Amuneal, Philadelphia, USA) which reduces the static magnetic flux by about a factor of 50 at the centre of the cryostat. This is necessary because the full strength Earth's magnetic field would impact the detector's performance and make it difficult or impossible to operate the SQUID-based preamplifiers used by SuperCDMS. The space between the inner lead layer and the cryostat is flushed with low-Rn air ($<$\,10$\;$Bq/m$^3$) by means of a purge-gas line going down to a diffuser at the bottom of the inner lead layer. An aluminum collar plate covers the gap between the cryostat and the lead shield (without mechanical contact between the cryostat and the shield) to ensure that a low air flow is sufficient to prevent the high-Rn air from the lab ($\sim$\,130$\;$Bq/m$^3$) from diffusing into that space.

\begin{figure}
\centering
  \includegraphics[width=\textwidth]{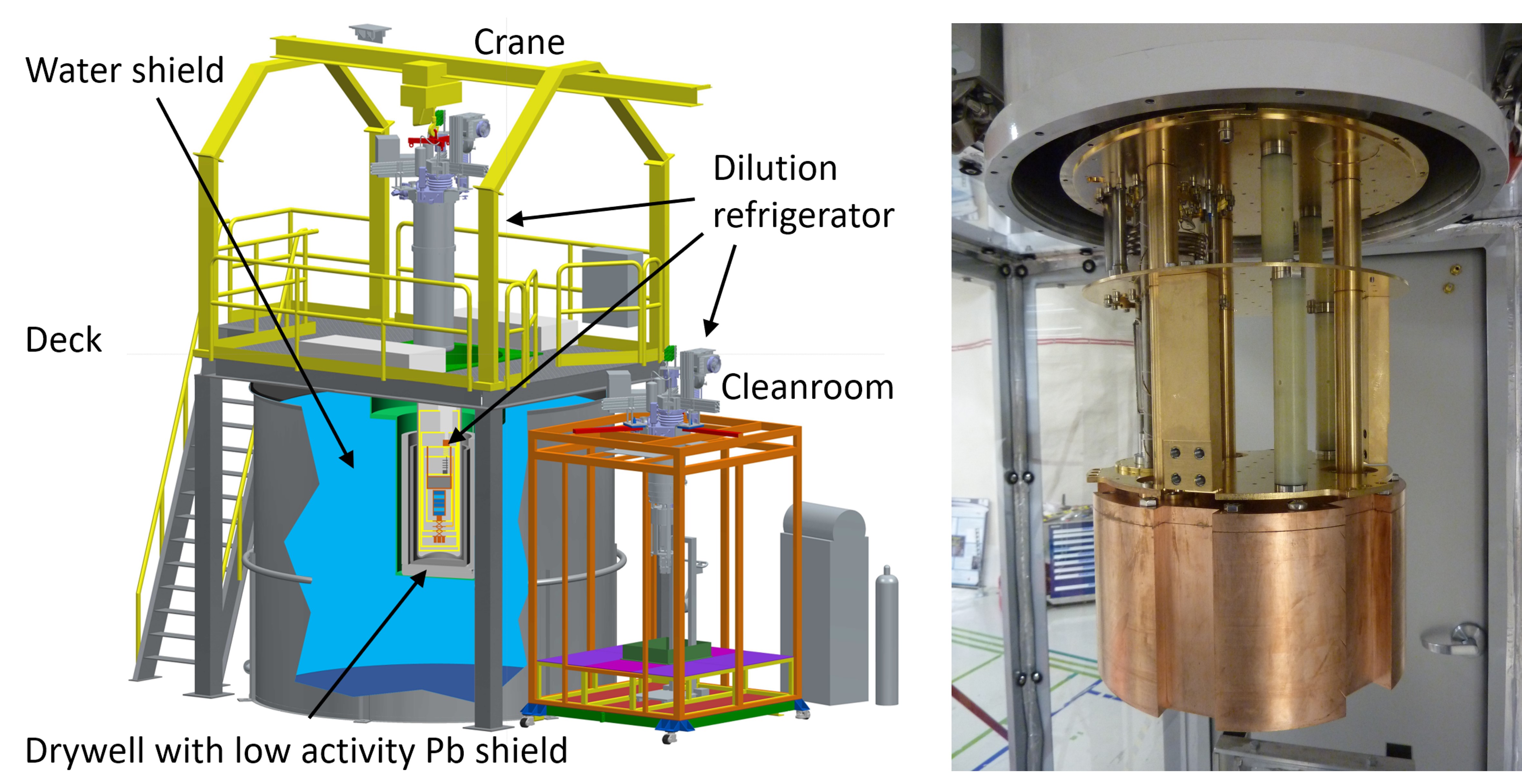} 
  \caption{Left: Layout of CUTE facility. While in operation, the dilution refrigerator hosting the payload is located inside the drywell at the center of the $\sim$\,3.5-m diameter water tank and surrounded by about 11$\;$cm of lead for shielding against environmental radiation. The deck structure holds a 20-cm thick polyethylene shield moderating neutrons coming from the top. It consists of two halves and is mounted on rails so it can be moved to the sides, giving access to the top of the cryostat. A monorail crane moves the cryostat between the drywell and the low-radon cleanroom for payload changes. Right: Internal lead shield sitting below the MC plate and thus also below all functional parts of the dilution refrigerator, blocking external gammas from the top as well as radiation from the dilution unit to the payload which would be mounted below (see Fig.~\ref{fig:cryo_1}). Mechanically and thermally it is attached to the Still stage of the refrigerator by three gold-plated copper rods that clear openings in the CP and MC plates of the fridge.}
  \label{fig:cute}
\end{figure}

The setup as described so far shields the experimental space inside the cryostat well from radiation from the floor and walls, but the presence of the refrigerator necessarily generates a big opening on the top. This opening is closed by a 20$\;$cm-thick polyethylene (PE) lid on the deck, mitigating neutrons from the top. The PE is encased in a stainless steel box and moves on rails to the side to provide access to the top of the refrigerator. Inside the cryostat, directly above the experimental volume (though thermally connected to the Still stage of the cryostat) is a 15$\;$cm-thick layer of lead encased in  copper to shield the experimental volume from gamma radiation from above, including from contaminants inside the dilution refrigerator. A sketch of the experimental setup and a picture of the internal lead shielding are shown in Fig.~\ref{fig:cute}. 

Most of the materials in use at CUTE were screened to assess their contamination levels. For the materials which were not screened, contamination levels of comparable materials from previous screening campaigns within the SuperCDMS experiment were considered. 

To estimate the background budget of the facility, extensive Geant4~\cite{Agostinelli:2003} Monte Carlo simulations were carried out. All the components of the facility were simulated considering all contaminants that were identified in the screening measurements. The gamma and neutron flux from the walls of the SNOLAB cavern are also simulated, considering the measured U, Th and \textsuperscript{40}K contamination. The simulations framework was set to generate gammas and neutrons from the bulks of the materials inside the facility and the surfaces of the SNOLAB cavern walls.

\begin{figure}
\centering
    \includegraphics[width=\textwidth]{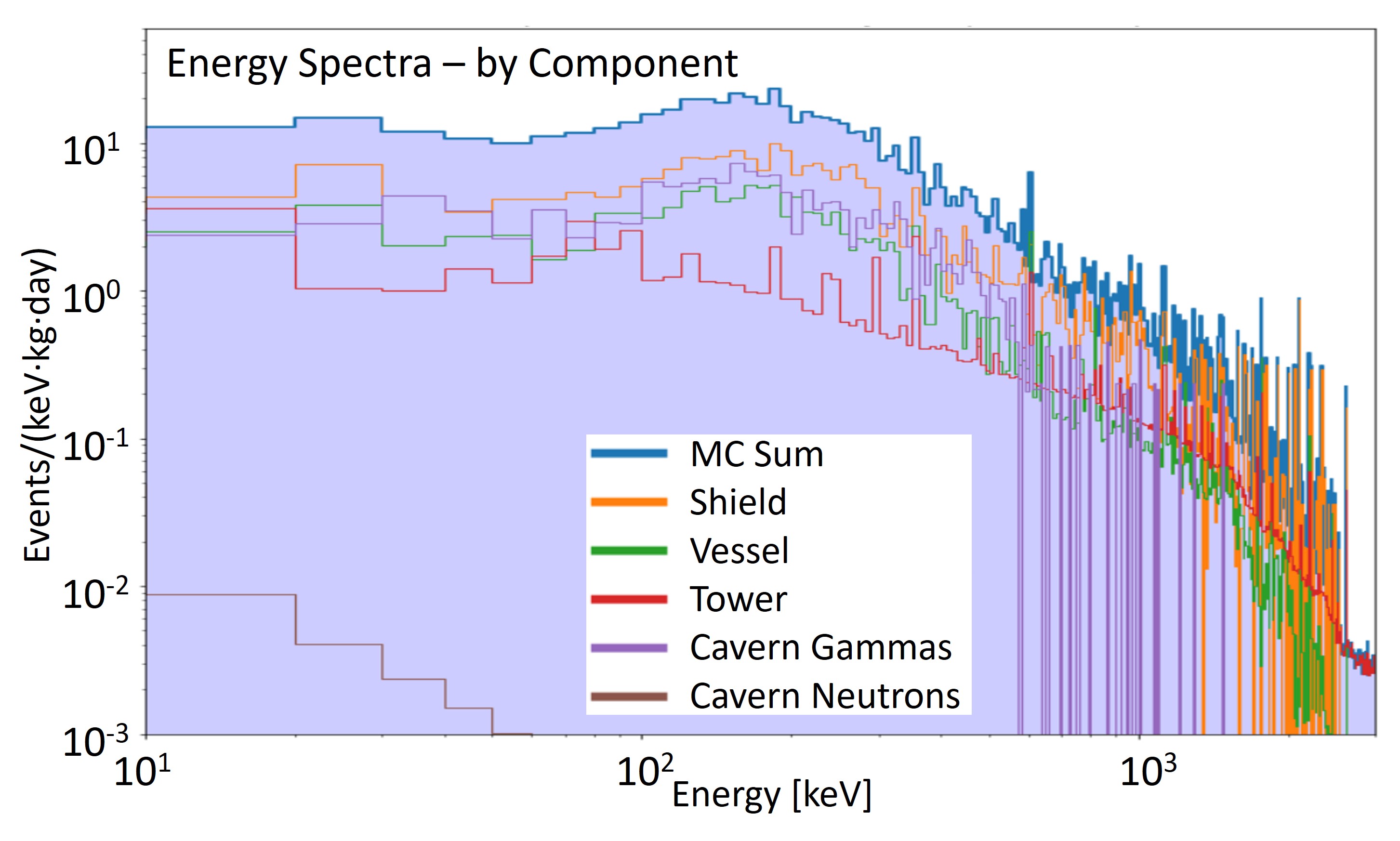}
  \caption{Energy spectra from Geant4 Monte Carlo simulations for different components of the background radiation in the CUTE facility for a 600-g SuperCDMS germanium detector. The sum of the contribution is shown in blue. ``Vessel'' refers to the cryostat components (including the internal shielding) while ``Shield'' includes all shielding components external to the cryostat. Note that most of the cryostat components had contamination levels below the sensitivity of the screening measurements so upper contamination limits from those measurements were used for the simulation. The ``Tower'' energy spectrum refers to the radioactive background induced by the detector target material itself and the sub-components of the detector tower structure. Nuclear recoils from radiogenic neutrons originating from the cavern wall contribute only a very small rate}
  \label{fig:sims}
\end{figure}

While the Geant4 Monte Carlo simulation propagates the radiation particles through the different components of the setup, Background EXplorer~\cite{BE} -- a tool originally developed by SuperCDMS -- handles normalization and conversion of simulated spectra into event rates. The energy spectra of events recorded in a 600-g Ge SuperCDMS germanium detector from the different simulated sources are shown in Fig.~\ref{fig:sims}. 

The sum of all components results in an event rate of 6.7$\pm$0.8$\;$events/keV/kg/day in the energy range from  1 to 1000$\;$keV (in reasonably good agreement with initial measurements) where about 10$\;$\% of the rate is contributed by the detector stack. The major contributors to the background budget are the gammas from the SNOLAB cavern ($\sim$\,30\%), the inner layer of the external lead shield ($\sim$\,20\%) and the stainless steel of the OVC ($\sim$\,13$\;$\%). The nuclear recoil rate is expected to be less than half an event/kg/day in the range from 1 to 50$\;$keV.

Most of the external gammas enter through the gaps between the external and internal lead shielding. If a lower gamma background is required for future measurements, an upgrade to the facility could improve the situation by adding additional gamma shielding to reduce those gaps, and replacing the highest contributors from the facility (inner layer of the external lead shield and OVC) with lower activity materials (lower activity lead for the shield and e.g.\ copper for the OVC).

\section{Calibration Systems}
\label{sec:calibration}
The low radiation environment provided by CUTE is not only of interest for testing cryogenic devices in the near absence of radiation, but also allows for dedicated tests of their response to radiation in a controlled way. For particle detectors this is important for calibration measurements, but may also be of relevance to other devices. One example is the impact of ionizing radiation on the coherence time and correlated error rates of superconducting qubits \cite{Oliver:2020, Cardani:2023}.

CUTE has two dedicated calibration systems, a gamma calibration system that was installed when the facility was first established, and a neutron calibration system which is presently in the process of being completed. In addition to these two systems that are both external to the cryostat, there is the possibility to install sources of radiation with low penetration inside the cryostat. An \textsuperscript{55}Fe X-ray source ($\sim$\,6$\;$keV) with an Al foil to generate X-ray fluorescence ($\sim$\,1.5$\;$keV) is presently available and has been used in CUTE. The use of other sources is possible, but they would need to undergo the approval process at SNOLAB.

\begin{figure}
    \centering
    \includegraphics[width=\textwidth]{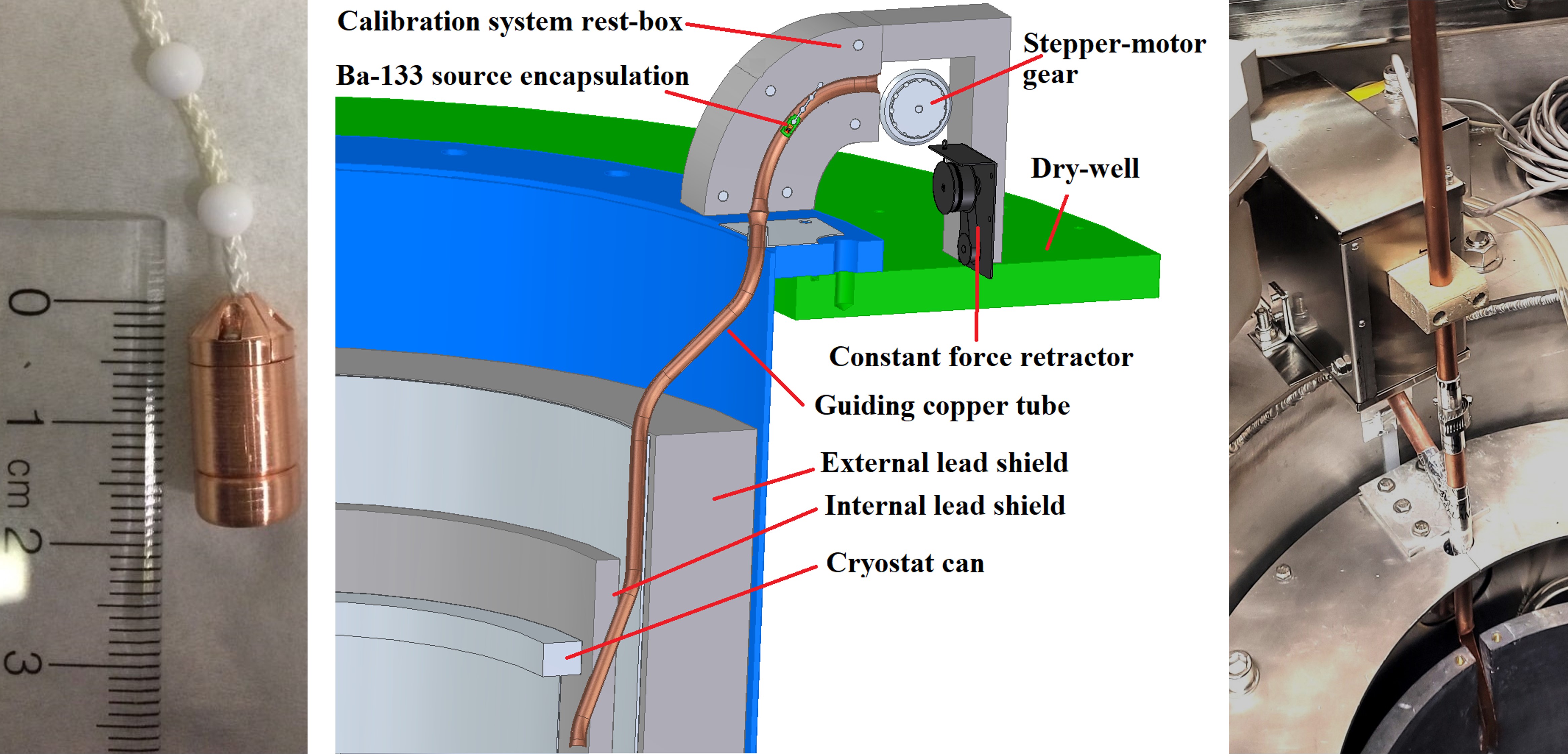}
    \caption{Left: Gamma calibration source (\textsuperscript{133}Ba, 37$\;$kBq nominal activity) with double encapsulation added to the manufactures encapsulation and attached to the beaded string. Middle: Gamma calibration system schematic. The source is stored in a lead housing located inside the drywell near the top of the cryostat. It is attached to a beaded string that is moved by a stepper motor (only its gear is shown) and spooled by a retractor. The copper tube guides the source into the space between the cryostat and its lead shielding. Right: As an upgrade, an additional guide tube was installed to allow the manual deployment of alternate sources into the same space.}
    \label{fig:gamma_calibration}
\end{figure}

The gamma calibration system is based on a \textsuperscript{133}Ba source with a design activity of 37$\;$kBq, doubly encapsulated in addition to the manufacturer's encapsulation (according to SNOLAB requirements) and attached to a beaded string (see left panel of Fig.~\ref{fig:gamma_calibration}). This string is attached to a constant-force retractor and its movement is controlled by a stepper motor. The system is installed inside the drywell, near the top of the cryostat. It includes a lead housing where the source is located when not in use. The source is pulled into the housing via a curved channel to ensure that there is no direct line of sight to the outside world when the source is in the storage location. With the $\geq$4$\;$cm of Pb surrounding the source in all directions, the radiation outside is negligible ($\ll$1$\;$nSv/h). A sketch of the calibration system set up is shown in the middle panel of Fig.~\ref{fig:gamma_calibration}.

When being deployed, the source is pulled by gravity, and after exiting the lead shielding it is guided by a copper tube past the top flange of the cryostat into the space between the cryostat and the lead shielding where it can be lowered as low as the bottom of the cryostat. When the source is fully retracted into the shielding, it activates a sensor referred to as {\it home sensor}, indicating that the sources is in the storage location. An IR reflection sensor package (IR LED and light sensor) is installed next to the chain, between the stepper motor and the lead shield where it detects the beads on the chain moving past to give feedback on the actual motion of the source.

As an upgrade to the original design, the copper guiding tube was modified to allow the deployment of alternate sources by hand as long as they fit into the tube (1/2 inch ID) and can be securely attached to a string (see right panel of Fig.~\ref{fig:gamma_calibration}; the system prior to this modification can also be seen in the top right of Fig.~\ref{fig:suspension}).

Neutron sources are of particular concern at SNOLAB as there are several experiments searching for nuclear recoils induced by dark matter particles which could potentially be mimicked by neutrons. Therefore, neutron sources must be stored in a way that ensures that their presence does not increase the neutron flux in those experiments at a measurable level. Additionally, each time a neutron source is moved through the lab, this action needs to be announced a week ahead of time, so the concerned experiments can take the presence of that source into account (or object if the experiment is in a critical phase). This severely restricts the flexibility when using neutron sources. 

Hence, the neutron source system for CUTE is designed such that during normal use the source (a \textsuperscript{252}Cf with a nominal activity of 37.5$\;$kBq) never leaves the shielding: it is stored within the CUTE water shielding tank, at the bottom near the edge of the tank, and when deployed it is located at the outside wall of the drywell, still inside the water tank, but close to the detectors with no or very little water in between the source and the detectors. The lead shielding inside the drywell is still in place. This diffuses the neutron flux, but does not drastically modify the energy spectrum. An option is built into the system to move the source away from the drywell in a controlled way, increasing the water layer between the source and the drywell from zero up to about 10$\;$cm to moderate the neutron spectrum (though reducing the total neutron flux at the same time). This is of interest if lower interaction energies are desired, and is also a powerful tool for the validation of Monte Carlo simulations of neutron interactions.

When in the storage location, the source is surrounded by gamma shielding which in turn is located inside a polyethylene (PE) box with about 30$\;$cm of PE all around except towards the bottom. The PE box ensures that at 50$\;$nSv/h, the dose rate outside the water tank is well below any level that would be of concern for personnel safety even if the water is removed from the tank. 

\begin{figure}
    \centering
    \includegraphics[width=0.8\textwidth]{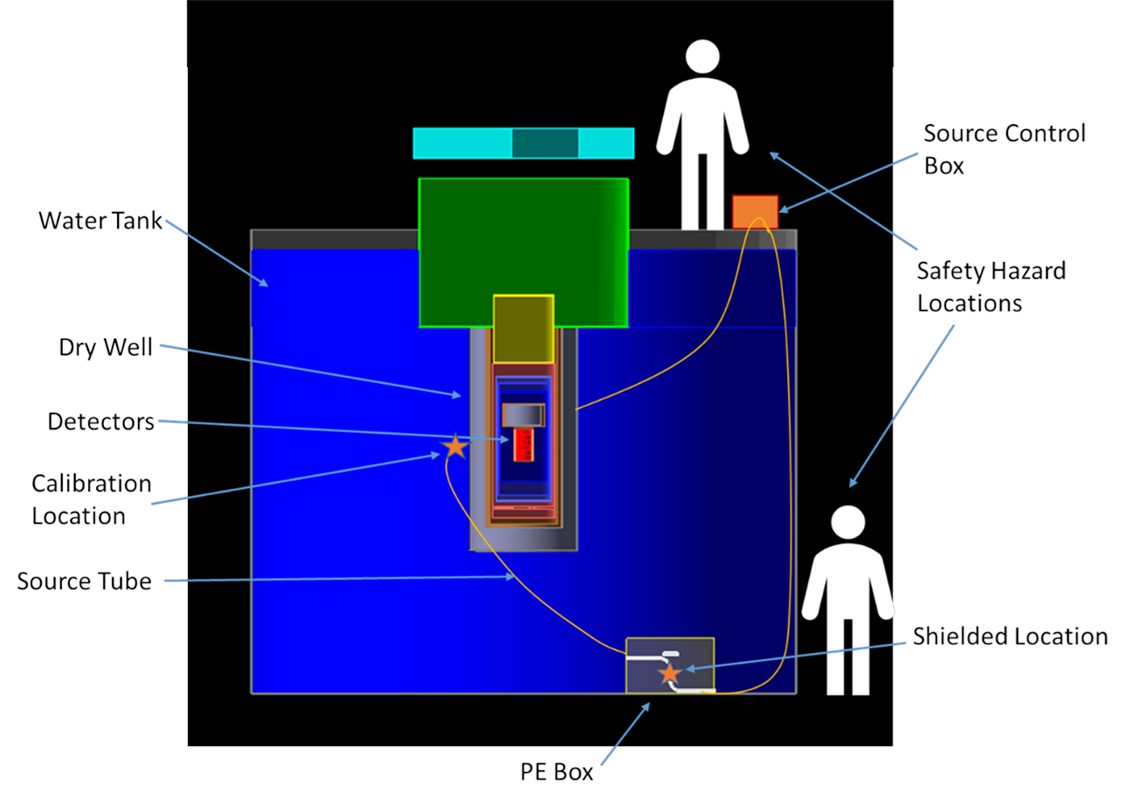}
    \caption{Schematic of the neutron calibration system from \cite{Corbett:2021}. The source is moved by a stepper motor (orange box on the top right) through a tube that is installed inside the water shielding tank. The main locations of the source (inside the storage box, bottom right, and in the calibration position, next to the drywell) are indicated by orange stars. Monte Carlo simulations have been performed to asses the radiation exposure to personnel in the two indicated locations, demonstrating that any exposure would be negligible.}
    \label{fig:neutron_calibration_schematic}
\end{figure}

Both sides of the source are attached to a beaded string (same type as used in the gamma calibration system). The string with the source forms a loop that moves inside the plastic tube through the water tank. The tube extends to above the lid of the tank where a stepper motor is located that controls the movement of the source. Special beads are attached to the chain at a defined distance away from either side of the source. If one of these beads is detected by a sensor inside the motor box, the source is in its storage location; if the other one is detected, the source is deployed near the drywell, the farthest away from the storage location the source can reach during normal operation. However, if necessary, the source can be brought all the way up into the motor box where it can be accessed and removed from the system. Figure \ref{fig:neutron_calibration_schematic} shows a schematic of the neutron calibration system.

\section{Payload Changes}
\label{sec:cleanroom}

In a low-background experiment, the most critical part that needs to be protected from contamination is the detector itself. This means the installation or removal of the payload in CUTE must happen under especially clean conditions. While SNOLAB is operated as a class-2000 cleanroom, extended exposure of the SuperCDMS detectors to the lab air would still lead to an accumulation of dust, contributing significantly to the total detector background. An even more important contribution to the background would come from the high radon concentration of typically 130$\;$Bq/m\textsuperscript{3} in the air at SNOLAB \cite{SNOLABhandbook}. To protect the payload from exposure to dust and radon, a dedicated cleanroom was installed next to the CUTE water tank. This cleanroom is supplied with low-radon air, either compressed air brought into the lab from the surface with a typical Rn concentration of $<$\,10$\;$Bq/m\textsuperscript{3}, or air from the SuperCDMS radon reduction facility with a Rn concentration of $\ll$\,1$\;$Bq/m\textsuperscript{3}, and is operated at a slight over-pressure. The air quality is constantly monitored by a NUVAP monitor \cite{nuvap}, recording among other parameters the particulate rates and the concentration of Rn. The cleanroom class is roughly 200, and for operation with surface compressed air, the radon concentration is at or below 10$\;$Bq/m\textsuperscript{3}. For safety reasons, there is also an oxygen monitor inside the cleanroom with the possibility to read the oxygen concentration from outside before entering.

A monorail crane is used to lift the cryostat from its operating position in the drywell to the cleanroom, where it is inserted through an opening in the roof. The centre section of the cleanroom roof has two sliding panels, each with a semi-circular cutout made to fit tightly around the top part of the OVC can; the two halves can be locked together. When the cryostat is not in the cleanroom, a plastic disk closes the hole. When the cryostat is in the cleanroom, the various cans can be removed and installed using a hand-operated lifting platform in the centre of the cleanroom, directly underneath the cryostat. When the cryostat is open, the lower sections of the cryostat cans sit on that platform, nested within each other. Plastic collars are used to prevent the inner cans from sliding into the outer ones thus keeping the flanges easily accessible for remounting. Cover plates are available to be placed on top of the cans, providing a working surface directly underneath the cryostat to facilitate the payload installation. Dedicated sets of cryostat and detector tools are located permanently in the cleanroom for convenience. A small movable table is available inside the cleanroom for preparing the payload.

Full cleanroom gear (cleanroom suit, double gloves, booties and hair nets) must be worn when entering the cleanroom to ensure the best cleanliness standards. In the absence of an airlock, the air quality (particulates and Rn concentration) is compromised when the door is opened to enter the room. It is therefore important to let the air conditions settle (characteristic time for this process is $\sim$\,10-20~minutes) before exposing any critical components. Given the small size of the cleanroom, only two people are allowed to work inside at any given time. A third crew member is often located outside to aid with the work planning, coordination, and helping the crew inside with operating procedures and logging of activities. Communication is facilitated by means of a phone line, video conferencing technology and the fact that the walls of the cleanroom are transparent for instantaneous full visual feedback.

\begin{figure}
\centering
  \includegraphics[width=\textwidth]{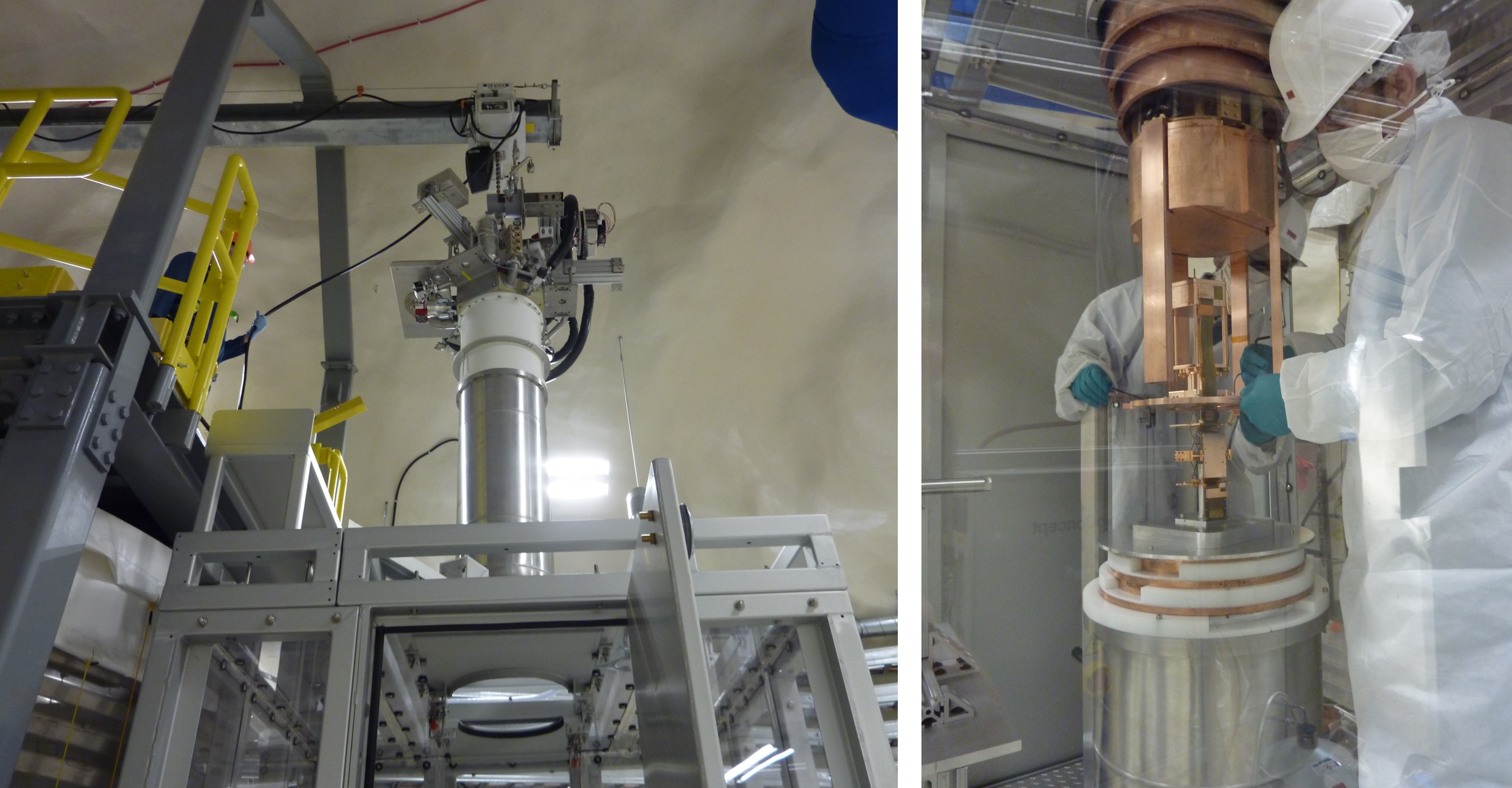} 
  \caption{Left: The Dilution refrigerator is being moved from the drywell into the cleanroom by the monorail crane. The plastic disk in the ceiling of the cleanroom has been removed and the sliding panels are open, ready for the cryostat to be lowered. Right: A SuperCDMS tower with a single detector about to be mounted to the refrigerator is resting on a cover plate atop the nested cans} \label{fig:payload}
\end{figure}

\section{Network and Electronics}
\label{sec:electronics}
The computer network for the facility is integrated into the SNOLAB network and is protected by a firewall. Access from the outside world is only possible by connecting to a virtual private network (VPN) which allows direct or indirect access to all network-enabled devices that are part of the facility. The present setup of the computer network makes use of two subnets: one is exclusive to the devices needed for the detector operation and readout and the other serves all other devices. A total of four computers are part of the network: one for the operation of the dilution refrigerator, one for all other slow-control activities, one for the data acquisition and one for data handling and transfer. In addition, CUTE has dedicated resources available at the SNOLAB surface facility, for data handling (receiving data from underground and sending them to partner institutions) and for some modest amount of data processing primarily for data quality control. 

CUTE's whole underground computing and network infrastructure is powered by an uninterruptable power supply (Eaton 9PX 6000 UPS). This UPS comes with an extension module with eighteen 120$\;$V power outlets, and allows for online oversight of the UPS performance and manual shutoff of individual outlets. In the current configuration, this UPS will provide power for about an hour in case of a power outage. As an upgrade to the facility, a second UPS was installed that powers the whole facility (Eaton 93E, 40$\;$kW). However, it can provide power only for about 10~minutes. In case of a power outage, this is usually enough to bridge the facility until SNOLAB's backup generator is able to restore power to the whole laboratory.

The slow-control system will be discussed in more detail in Sec.~\ref{sec:slow-control}. The rest of this section is dedicated to a short description of the SuperCDMS-specific electronics and detector readout solutions. 

As discussed in Sec.~\ref{sec:design}, the SuperCDMS detectors are mounted on a structure that includes the wiring between the three lowest temperature stages as well as key components of the first-stage amplifier electronics. For each detector installed, a cable with 100 individual conductors (wire loom with 50 twisted pairs) is attached to the tower-wiring and makes the connection to the 4K stage of the cryostat; this cable is superconducting to minimize the conductive thermal load on the Still stage. From here another cable carries the signals to the room-temperature vacuum interface at the top of the cryostat. The vacuum feedthrough is achieved by a custom-designed printed circuit board (the Vacuum Interface Board, VIB) sandwiched between an ISO160 stainless steel flange. On the outside, custom-designed Detector Control and Readout Cards (DCRCs) attach directly to the VIB. These DCRCs hold the complete control and readout electronics including the signal digitization. They communicate directly with the data acquisition computer via an Ethernet connection and are powered via a 48-V Power-over-Ethernet (PoE) power supply. High Voltage (HV) for the detector bias is provided to the boards through a dedicated connector on the VIB. The 24-port PoE as well as the HV power supply (eight positive and eight negative HV outputs) are identical to those used in SuperCDMS and will be made available as backup for the respective components in the experiment if needed, but otherwise are part of the CUTE facility and will in the future be available for other users of CUTE. Both devices can be addressed and controlled remotely.

\section{Slow-control system}
\label{sec:slow-control}
A crucial aspect of the CUTE facility is its slow-control system which encompasses the monitoring and control for the cryogenic subsystems, the suspension system, the calibration source deployment systems and all other devices and sensors at the facility other than the payload and its readout system. All of the available facility data are recorded and stored in a MySQL database on the slow-control computer. 

The control and logging software for the dilution refrigerator and its auxiliary systems was developed and provided by the manufacturer of the dilution refrigerator, CryoConcept, and is installed on the fridge control PC. This software provides functionality to start and stop the pumps and compressors, open  and close valves, control thermometry and heater settings, read out and log the thermometry and pressure gauges, and set parameters for automated tasks like `cool down' or `warm up'. While it provides the necessary functionality for normal fridge operations, it leaves room for improvements with regards to the monitoring and control of the compressor that drives the pulse tube cooler. In particular, the CryoConcept software cannot reset the compressor which is necessary after the occurance of certain types of errors (e.g.\ when the cooling water temperature is out of range) or after a power outage. It also does not log the information from the compressor, such as water, oil and helium temperatures. Python scripts were developed to make those functionalities available.

The fridge control PC also runs a program monitoring the liquid nitrogen (LN) cold trap: a 30-L LN Dewar containing the trap sits on a scale which is continually read out by a Python script. The weight can then be translated into the amount of LN in the Dewar. The system also includes a low-pressure 100-L LN Dewar equipped with a heater used to build up pressure for transferring LN to the cold-trap Dewar when needed.

The Peltier cooler for the turbo pump is operated by a hardware controller to maintain its temperature. A Python script was developed to interface with the controller to start and stop the cooler, set the temperature and log the temperature and the output power of the controller.

The suspension system and the calibration systems are both operated through AVR microcontrollers. To communicate with the AVRs, a server written in Node.JS is running on the slow-control computer. Each AVR microcontroller is flashed with a C code that handles the driving of the stepper motors, the I/O of the digital pins, the measurements of the ADCs, and the USB communications. The actual logic used by the suspension system and calibration systems is implemented in the Node server, and the microcontroller code only handles the low level hardware control. This provides an additional layer of flexibility by being able to easily add new functionality to the system without having to reprogram the AVR which would involve modifying the (generally) more complicated C code. The Node server handles creating the USB connection to the microcontrollers, and allows clients to connect to and communicate with the server via a websocket interface.

The logic for the suspension system control software is based on continually reading out the optical sensors via an ADAM-6017 module. A nominal position for the suspension system is set by the user; while the original design anticipated this to be in the centre of the 2$\;$mm-vertical range of the suspension system (that is at 1.0$\;$mm), it can be set to any value between 0 and 2$\;$mm. If a sensor reading deviates from the nominal position by more than a predefined amount (presently set to 0.04$\;$mm which was found to be a good compromise between stability of operation and adjustment frequency), the respective stepper motor is driven in the appropriate direction to adjust the damper position and thus bring it back to within range. A safety feature stops the automation if an extended adjustment by the stepper motor does not lead to any measurable change in the position. In addition to the automated position control, each stepper motor can be individually controlled by the user to move up or down with an adjustable speed.  

The compression of the dampers as determined by the Hall effect sensor together with the optical sensor is converted to an equivalent mass that should be added or removed from each damper to bring it to its nominal operating point. This is mostly relevant when a new payload with a significantly different mass has been installed. Weights can be added to or removed from the suspension frame to bring the compression into the desired range.

The motor that deploys and retracts the calibration source is  also controlled by the Node server, and users can send the sources to specific positions via the web interface. The position is reached by driving for a specific number of steps, as the travel of the source per motor step is known. The feedback sensors (see Sec.~\ref{sec:calibration}) are used to confirm that the source is moving as intended (optical sensor) or to confirm it is in the housing (home sensor). This is useful after a power outage (in which case the information on the present position can get lost) or as a quick feedback if the source is not moving as intended.

\begin{figure}
    \centering
    \includegraphics[width=\textwidth]{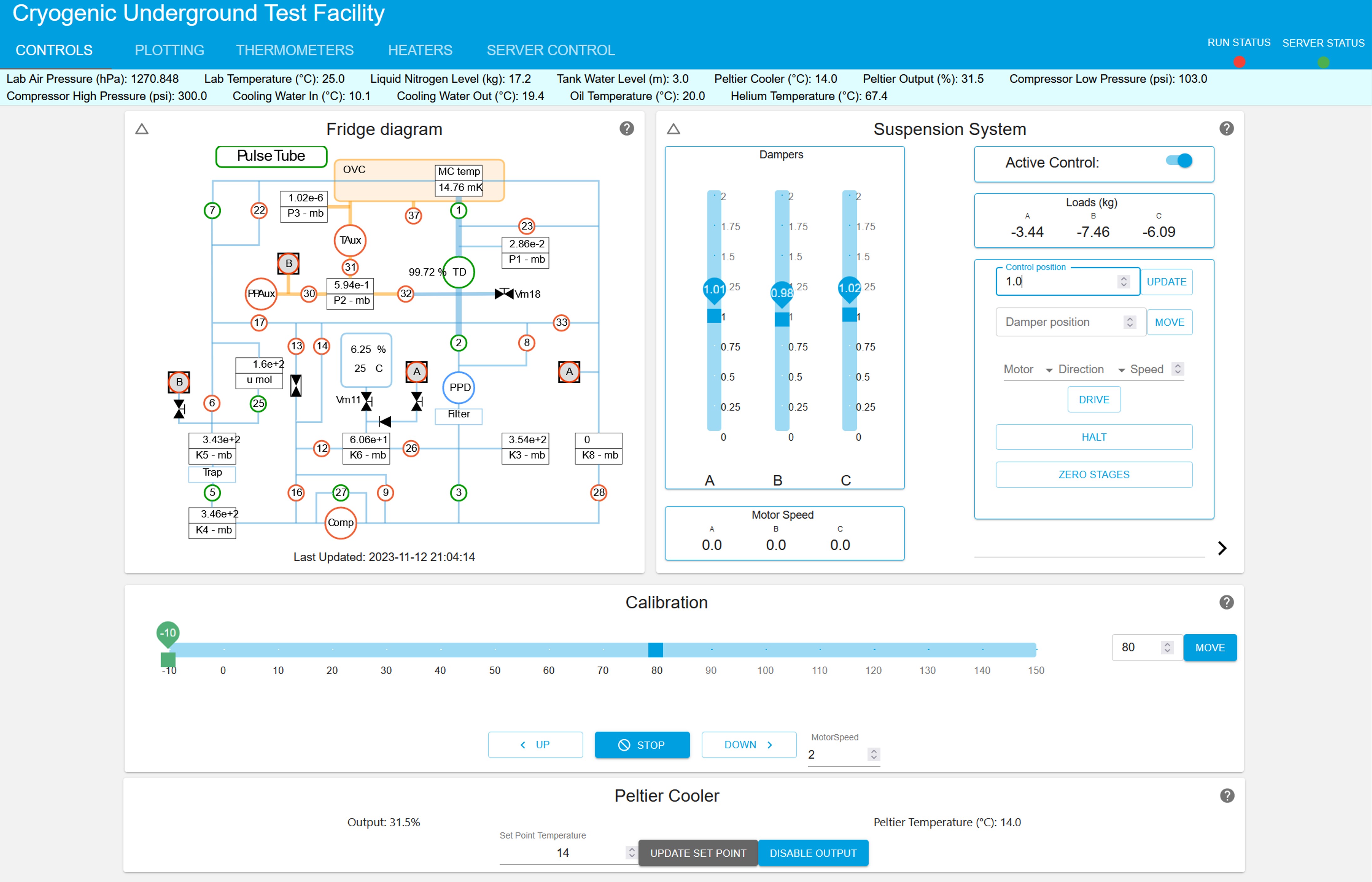}
    \caption{Screen-shot of the main slow-control webpage. In the top half are the fridge monitoring panel (left) and the panel for the monitoring and control of the suspension system (right). Active elements in the fridge monitoring panel (pumps, valves) are green, while inactive ones are red. The gamma calibration source is controlled through the first horizontal panel in the lower half; it indicates that the source is located inside the shielding box (``$-$10$\;$cm"; 0 is just outside the lead shielding and level with the top rim of the lower part of the drywell). The lower horizontal bar controls the Peltier cooler. The cyan-colored banner above the panels shows various facility parameters (including lab air temperature and pressure). The blue top-banner allows the user to choose the alternate tabs for plotting and data downloading, control of the fridge thermometers and heaters, and an interface to easily check the status of (and if necessary restart) the servers for the suspension and calibration systems, the Peltier cooler and the compressor monitoring.}
    \label{fig:slow_control}
\end{figure}

A webpage was designed and built that brings much of the monitoring and control for the various subsystems into one convenient location. The status of the cryogenic systems, suspension system, the Ba calibration system and the Peltier cooler are prominently featured, and key parameters from across the facility are displayed in a ribbon along the top of the page. The suspension system, the calibration system and the Peltier cooler can also be controlled through this webpage. In addition, it includes a feature for plotting and downloading data from the database for easy monitoring of the performance of the facility over time. Most of the functionality of the fridge control software was deliberately not included in this webpage: while the slow-control webpage is a one-stop-shop for all users of the facility, the operation of the dilution refrigerator is limited to a small number of experts. Figure \ref{fig:slow_control} shows a screen shot of the web page.

\section{Remote Operation}
Accessing SNOLAB comes with considerable extra effort compared to most other work places, due to its location 2$\;$km underground in an active mine and the cleanliness requirements in the lab. In addition, there is no regular access during weekends and holidays, including extended no-access periods during the winter holidays and typically for several weeks during the summer for maintenance work by the mining company. Therefore it is important to be able to remotely monitor and control various aspects of the facility and experiment. Significant effort was made to ensure that CUTE can operate without operators present at the lab for an extended period of time. 

The fridge control software operates on a remotely accessible PC and includes all functionality that is required for operating the dilution refrigerator. With the above mentioned custom additions, this also includes a restart after a loss of power. With the nitrogen refill system, that can also be operated remotely, the dilution refrigerator can operate for up to about two months without personnel underground. As discussed in Sec.~\ref{sec:slow-control}, all relevant systems are remotely controllable and all facility data are accessible through the database. In addition to the aforementioned systems, a remote controlled power distribution unit with five outlets is located at the deck, for connecting and switching auxiliary devices on and off. Finally, the PoE and the HV power supplies are fully remotely controllable.

If all services are available, the period for complete remote operation is limited by the LN supply. Power interruptions can be bridged for about 10~minutes (see Sec.~\ref{sec:electronics}). If the facility loses power, it can be recovered remotely, provided the other systems are operating. However, depending on the length of the outage the cryostat may have warmed up significantly in which case the recovery would take an extended period and may consume significantly more LN than in steady state operations. For compressed air, required for switching the valves of the dilution refrigerator, there is presently a small buffer tank. In steady state operation when no valves are changing state, this lasts for an extended period, but runs quickly out when valves are operated. The backup is sufficient to collect and secure the helium mixture in case an extended outage is expected. For cooling water, CUTE is connected to the low-pressure cooling water loop installed for SuperCDMS. This system presently has no backup in case of failure (either of the low-pressure loop itself, or the primary high-pressure cooling loop operated by SNOLAB). In case the cooling water fails, the compressor for the pulse tube cooler will stop operating after roughly a minute or two. SuperCDMS is exploring options to ensure the long-term stability of the system.

\section{Use of CUTE}
The purpose of CUTE is to provide an environment to test cryogenic devices that are known or suspected to be affected by radiation and vibrations. Prime examples are cryogenic detectors for rare event searches such as dark matter direct detection, neutrinoless double beta-decay, or rare nuclear decays. Depending on the expected event rate, the size and background of the facility may also be sufficient to carry out actual rare event searches. 

As discussed earlier, the original motivation for building this facility was the testing of the new SuperCDMS detectors under low-background conditions. The facility conditions are such that operating one of the SuperCDMS HV detector towers for a few months could push the sensitivity of SuperCDMS by several orders of magnitude, though would fall short of the goals of SuperCDMS SNOLAB by one to two orders of magnitude due to the higher background and the smaller payload.

Besides cryogenic particle detectors there are other devices that may benefit from the special environment provided by CUTE, such as cryogenic devices being developed in the fields of quantum information and sensing. These emerging fields have the potential to realize major advancements in a number of areas. One of the leading platforms in this area are superconducting quantum devices, due to their modular design and the fact that they can be easily fabricated using techniques developed by the semiconductor industry. Recently, it was shown that ionizing radiation could generate excess quasiparticles in superconducting qubits, thereby degrading their performance in a way that would not be easily handled by error correcting codes \cite{Oliver:2020, Cardani:2023}. Operating such devices in a low-background environment would reduce the error rate due to particle interactions significantly and open the door to study (and potentially resolve) more fundamental limitations.

Interested parties can submit proposals for the use of CUTE for particular measurements\footnote{Send proposals via e-mail to {\tt cute$\_$proposals@snolab.ca} or contact SNOLAB management}. Once a proposal has been accepted, the CUTE team will work with users on the installation plan and help with all facility interfaces. While the users are generally responsible for providing all necessary hardware for the device installation and readout, including cabling and readout electronics, some basic equipment will be available and the aim is to constantly improve the facility to provide more support. For example, a microwave setup consisting of cryogenic coaxial cables, attenuators, filters, and amplifiers is planned to be installed for a specific experiment, but is likely to stay at CUTE and be available afterwards for other users.

\section{Summary and Conclusion}
CUTE is an underground facility at SNOLAB, built for the purpose of testing cryogenic detectors and other devices in a low-background environment. The underground location together with a composite shielding of water, low-activity lead and polyethylene drastically reduce the flux of cosmogenic and radiogenic radiation. A layer of magnetic shielding reduces the Earth's magnetic field by about a factor of 50 at the location of the payload and the mechanical decoupling of the cryostat from the upper part of the dilution refrigerator (which includes the pulse tube cooler cold head) together with the cryostat's suspension system drastically reduces the level of vibrations which otherwise might negatively impact the performance of the devices being operated in CUTE. The CUTE facility currently includes a \textsuperscript{133}Ba source delivery system that can be used for detector calibrations and other studies of the impact of gamma radiation on devices, and a \textsuperscript{252}Cf neutron source delivery system is planned to be installed and commissioned soon; certain other sources may be made available upon request. Overall, the CUTE facility provides an excellent environment for cryogenic experiments that require low levels of radiation. While the primary motivation for the facility was to test SuperCDMS SNOLAB detectors in an environment with a similar background level as the main experiment, once the testing of SuperCDMS detectors in CUTE is complete, other experiments may apply to use the facility on a proposal basis. 

\section{Acknowledgments}
The construction of CUTE was made possible through the provision of funds by Dr.\ Gilles Gerbier (Queen's University) from his Canada Excellence Research Chair grant and generous support from SNOLAB. Additional funds were provided by Queen's University through a Special Research Grant held by Dr.\ Wolfgang Rau. The operation and a variety of upgrades were funded largely by the Natural Sciences and Engineering Research Council of Canada (SuperCDMS project grant and an individual grant held by Dr.\ Jeter Hall), the Arthur B.\ McDonald Canadian Astroparticle Physics Research Institute, through support for research groups at TRIUMF, the University of Toronto,  and the University of British Columbia as well as through engineering and technical support, and by the University of Toronto through a Special Research Grant held by Dr.\ Ziqing Hong. A contribution to the \textsuperscript{3}He-contingent was made by Dr.\ Matt Pyle (University of California, Berkeley). During the construction, the technical and engineering support at Queen's University was invaluable; we thank in particular Chuck Hearnes, Robert Gagnon and Phil Harvey for their contributions. We would like to thank SNOLAB and its staff for support through underground space, logistical and technical services. SNOLAB operations are supported by the Canada Foundation for Innovation and the Province of Ontario, with underground access provided by Vale Canada Limited at the Creighton mine site. Particular thanks for on-site support go to Jasmine Gauthier and Jos\'e Marco Olivares. A lot of students contributed during their internship at SNOLAB to the Monte Carlo simulations and CUTE detector payload changes; we especially thank Melissa Baiocchi, Scarlett Gauthier, Sabrina Cheng, Jordan Ducatel and Alexander Pleava for their strong commitment. We also thank the group of students from the British Columbia Institute of Technology, Sean Green, Karel Chanivecky Garcia, Gurden Singh Angi and Johnathon Gordon Scott who helped develop the slow-control webpage, and all other undergraduate Summer and Co-op students who were involved in the project over the years. Last but not least we thank the SuperCDMS Collaboration: for validating the performance of the facility, the operation of SuperCDMS devices was invaluable. We thank SuperCDMS for providing these devices as well as personnel to operate them and assess their performance.

\bibliographystyle{unsrt}
\bibliography{biblio}

\end{document}